\documentclass{aa}
\usepackage[applemac]{inputenc}
\usepackage{natbib}
\usepackage{txfonts}
\usepackage{graphicx}
\usepackage{color}
\bibpunct{(}{)}{;}{a}{}{,}

\begin{document}

\title{Radial molecular abundances and gas cooling in starless cores}
\author{O. Sipilä}
\institute{Department of Physics, PO Box 64, 00014 University of Helsinki, Finland\\
e-mail: \texttt{olli.sipila@helsinki.fi}
}

\date{Received / Accepted}

\abstract
{}
{We aim to simulate radial profiles of molecular abundances and the gas temperature in cold and heavily shielded starless cores by combining chemical and radiative transfer models. Attention is also given to the time-evolution of both the molecular abundances and the gas temperature.}
{A determination of the dust temperature in a modified Bonnor-Ebert sphere is used to calculate initial radial molecular abundance profiles. The abundances of selected cooling molecules corresponding to two different core ages are then extracted to determine the gas temperature at two time steps. The calculation is repeated in an iterative process yielding molecular abundances consistent with the gas temperature. Line emission profiles for selected substances are calculated using simulated abundance profiles.}
{The gas temperature is a function of time; the gas heats up as the core gets older because the cooling molecules are depleted onto grain surfaces. The change in gas temperature associated with depletion is of the order of 1\,K. The contributions of the various cooling molecules to the total cooling power change with time, but the main cooling molecule at all times, in the range of environments studied here, is CO. Radial chemical abundance profiles are non-trivial: different species present varying degrees of depletion and in some cases inward-increasing abundances profiles, even at $t > 10^5$ years. Line emission simulations indicate that cores of different ages can present significantly different line emission profiles, depending on the tracer species considered.}
{Chemical abundances and the associated line cooling power change as a function of time. Most chemical species are depleted onto grain surfaces at densities exceeding $\sim 10^5$\,cm$^{-3}$. Notable exceptions are $\rm NH_3$ and $\rm N_2H^+$; the latter is largely undepleted even at $n_{\rm H} \sim 10^6$\,cm$^{-3}$. On the other hand, chemical abundances are not significantly developed in regions of low gas density even at $t \sim 10^5$ years, revealed by inward-increasing abundance gradients. Except in high-density regions where the gas-dust coupling is significant, the gas temperature can be significantly different from the dust temperature. This may have implications on core stability. Owing to the potentially large changes in line emission profiles induced by the evolving chemical abundance gradients, our models support the idea that observed line emission profiles can, to some extent, be used to constrain the ages of starless cores.}


\maketitle

\section{Introduction}

Low-mass stars are born out of gravitationally bound concentrations of cold gas, the so-called starless cores. The very low temperature ($\sim 10$\,K) of these objects allows molecular depletion onto grain surfaces to proceed efficiently, particularly in the dense central areas of the cores. Indeed, molecular depletion has been observed toward many starless cores \citep[e.g.][]{Willacy98, Caselli99, Tafalla02, Jorgensen04}. Because depletion is more efficient at higher densities, one expects the abundances of most molecules to fall toward the core center. Common molecules such as CO or $\rm NH_3$ may however have (relatively) low abundances in young cores with low gas density, since in these conditions chemistry proceeds slowly. Therefore one can qualitatively expect the abundances of most species to peak at midrange ($\sim 10^3 - 10^4$\,cm$^{-3}$) densities, when studying regions where the gas density varies by multiple orders of magnitude.

It has been observed and suggested by theoretical models that in starless cores the gas temperature is generally different than the dust temperature \citep[e.g.][]{Galli02, Young04, Keto05, Bergin06}. This is because at low gas densities (corresponding to the outer parts of starless cores) the gas-dust coupling is weak so that the gas cools efficiently through line radiation. Starless cores are supported (mainly) by thermal pressure; knowledge of the gas temperature is thus imperative to understanding the stability of the cores and hence the initial conditions of low-mass star formation.

Because the line cooling power is determined by molecular abundances and the abundances themselves are time-dependent quantities, it is of interest to study how much the gas temperature can change as a function of time and to what extent possible temporal changes in gas temperature can feed back into the chemistry. Also, because observational data of starless cores is gathered through observations of line emission radiation, predictions of radial molecular abundances, and the associated line emission, as a function of time are useful when interpreting observations. In this paper, we combine chemical and radiative transfer calculations to study both the temporal and radial dependences of chemical abundances and the gas temperature in starless cores.

The paper is structured as follows. In Sect.\,\ref{s:modeling}, we introduce the physical and chemical models and discuss the calculations and the various assumptions made. Section \ref{s:results} presents the results of our calculations. In Sect.\,\ref{s:discussion} we discuss our results and in Sect.\,\ref{s:conclusions} we present our conclusions.

\section{Modeling} \label{s:modeling}

In this section we give a detailed account of the physical and chemical models used. We also discuss the determination of the gas temperature and the spectral line simulations.

\subsection{Physical model}

The physical core model is the modified Bonnor--Ebert sphere \citep[hereafter MBES;][]{Galli02, Keto05, Sipila11}. The MBES is a gas sphere in hydrostatic equilibrium bounded by external pressure. In contrast to the Bonnor--Ebert sphere \citep{Bonnor56} which is isothermal, the MBES presents a temperature structure that can be self--consistently calculated \citep{Sipila11}.

In this paper, we consider three different MBESs with masses 0.25\,${M_{\sun}}$, 1.0\,${M_{\sun}}$ and 5.0\,${M_{\sun}}$. We chose to study several cores with different masses because of their varying physical properties; low mass MBESs are dense and small, while high mass MBESs are rather diffuse and more extended \citep{Sipila11}. From the chemical point of view this means that we expect to see very different radial profiles for species such as CO in the different cores, because molecular depletion is more efficient at higher densities. In all cases, we assume $\xi_1 = 6.0$ for the non--dimensional radius of the MBES, so that the model cores should be stable \citep[see][where the non-dimensional radius $\xi_1$ is defined]{Sipila11}. We assume the \citet{OH94} thin ice mantle model for the optical properties of the dust and that the cores are embedded in a larger parent cloud with visual extinction corresponding to $\rm A_V = 10$ at the core edge -- with these choices, the model cores discussed here correspond to the ``type 1'' cores of \citet{Sipila11}. The physical parameters of the model MBESs are presented in Table\,\ref{tab1}; the three different cores are denominated ``A'', ``B'' and ``C''. The (dust) temperature structures of the cores are shown in Fig.\,\ref{fig:coolabu}.

\begin{table}
\caption{Physical parameters of the MBESs considered in this paper. From left to right, the columns show the model denomination, core mass, core radius, total hydrogen density in the center shell of the model and the total hydrogen density in the outermost shell of the model, respectively.}
\centering
\begin{tabular}{c c c c c}
\hline \hline 
Model & $M$\,[$M_\odot$] & $R_{\rm out}$\,[AU] & $n_{\rm H}^{\rm c}$\,[cm$^{-3}$]  & $n_{\rm H}^{\rm out}$\,[cm$^{-3}$]   \\ \hline
A & 0.25  & 3700 & $1.85\times10^6$ & $2.19\times10^5$ \\
B & 1.0  & 12600 & $1.82\times10^5$ & $1.95\times10^4$ \\
C & 5.0  & 55000 & $1.09\times10^4$ & $1.02\times10^3$ \\ \hline
\end{tabular}
\label{tab1}
\end{table} 

\subsection{Chemical model} \label{ss:chemmodel}

The chemical model program used in this paper utilizes the rate--equation method. The code is an updated version of the one used in \citet{Sipila10}, expanded to include adsorption, desorption and reactions on the surfaces of grains. The gas phase chemical reactions are adopted from the OSU reaction file {\tt osu\_03\_2008}\footnote{See {\tt  http://www.physics.ohio-state.edu/$\sim$eric/}}. We assume spherical grains with $a_{\rm g} = 0.1\,\mu \rm m$. The cosmic ray ionization rate is set to  $\zeta = 1.3\times10^{-17}$\,s$^{-1}$. The initial gas phase abundances and the surface reaction set (i.e. activation energies for select reactions) are adopted from \citet{Semenov10}. The adopted initial abundances are atomic (with the exception of hydrogen which is practically totally in $\rm H_2$) -- the influence of this assumption on the results of this paper is discussed in Sect.\,\ref{ss:molabus}. We assume that the total amount of matter on grain surfaces is initially zero, i.e. that the grains are bare. We note that this is a source of slight discrepancy with the dust temperature calculation, since the \citet{OH94} dust model assumes grains coated in ice.

The rate coefficient for adsorption of species $i$ onto grains is $k_i^{\rm ads} = v_i \, \sigma \, S_i$ [$\rm cm^3\,s^{-1}$], where $v_i = \sqrt{8k_BT_{\rm gas} / \pi m_i}$ is the thermal speed of species $i$, $\sigma = \pi a_{\rm g}^2$ is the grain cross section and $S_i$ is a sticking coefficient, assumed here to be unity for all neutral species (we assume that ions do not stick).

The adopted desorption process is cosmic ray desorption as described in \citet{HH93}, i.e. the rate coefficient for cosmic ray desorption is $k_{\rm CR} = f\,k_{\rm TD}(70\,\rm K)$, where $k_{\rm TD}$ is the thermal desorption coefficient at 70\,K and $f$ is an efficiency factor \citep[we assume $f = 3.16\times10^{-19}$ corresponding to the value of $\zeta$ adopted here;][]{HH93}. We do not include thermal desorption as a separate process because the core temperatures are low ($\leq$\,10\,K) -- in these conditions, thermal desorption is negligible compared to cosmic ray induced desorption. Also, the present model does not include photodesorption.

The rate coefficients of reactions on grain surfaces are calculated internally in the chemical model program, following the formalism of \citet{HHL92}. That is, the rate coefficient for a reaction between species $i$ and $j$ on the surface of a grain is calculated as
\begin{equation}
k_{ij} = \alpha \, \kappa_{ij} \, ( R_{\rm diff}^i + R_{\rm diff}^j ) / n_{\rm d} \, ,
\end{equation}
where $\alpha$ is the branching ratio of the reaction and $n_{\rm d}$ is the number density of dust grains. $\kappa_{ij}$ is an efficiency factor, assumed to be unity for exothermic reactions without activation energy. For endothermic reactions or exothermic reactions with activation energy, $\kappa_{ij} = \exp(-E_{\rm a} / T_{\rm dust})$ where $E_{\rm a}$ is the activation energy. The diffusion rate $R_{\rm diff}$ is calculated as
\begin{equation}
R_{\rm diff} = {\nu_0 \over N_s} \exp (-E_{\rm diff} / k_{\rm B} T_{\rm d} ) \, ,
\end{equation}
where $\nu_0 = \sqrt{2 n_s k_{\rm B} E_{\rm b} / \pi^2 m}$ is a characteristic vibration frequency, $N_s$ is the number of adsorption sites per grain and $n_s$ is the surface density of adsorption sites per grain. We assume $n_s = 1.5\times10^{15}$\,cm$^{-2}$, which leads to $N_s = 1.885\times10^6$ assuming spherical grains with $a_{\rm g} = 0.1\,\mu$m. We assume $E_{\rm diff} = 0.77\,E_{\rm b}$ for the diffusion energies; most of the binding energies $E_{\rm b}$ are adopted from \citet{Garrod06}, but there are two exceptions. For $\rm H_2$ we have used a value $E_{\rm b} = 100$\,K; as pointed out by \citet{Garrod11}, the use of binding energies appropriate to a surface covered in water ice can lead to unphysical $\rm H_2$ ice abundances. \citet{Garrod11} corrected for this by allowing the binding energies of all species to change as the amount of $\rm H_2$ present on grain surfaces changes. Artificially lowering the $\rm H_2$ binding energy to $E_{\rm b} = 100$\,K lowers the steady-state $\rm H_2$ ice abundance to about 10\,\% of the $\rm H_2O$ ice abundance, depending on the total hydrogen density. Our approach is rather arbitrary, but avoids the problem of increasing the number of equations to solve -- decreasing the binding energy of $\rm H_2$ did not in our tests have a noticeable effect on the abundances of other species. For the binding energy of atomic oxygen we have used a value of 1390\, K \citep{Bergeron08, Cazaux11}. Increasing the binding energy of O from 800\, K \citep{Garrod06} to 1390\, K changes greatly the behavior of $\rm O_2$ at high densities; using the lower value leads to radial $\rm O_2$ abundances of $\sim 10^{-6}$ even in high density regions, inconsistent with observations \citep{Goldsmith00, Liseau12}. The higher binding energy value significantly decreases the radial $\rm O_2$ abundance at high densities (see Sect.\,\ref{s:results}, where the radial $\rm O_2$ abundances are further discussed).

The effect of quantum tunneling on grain surfaces has been subject to some controversy. \citet{Katz99} argued that quantum tunneling of hydrogen on amorphous surfaces would be inefficient. On the other hand, recent results of \citet{Goumans10} suggest that quantum tunneling dominates the $\rm CO + O \rightarrow CO_2$ reaction at low temperatures (this reaction is however very slow at typical starless core temperatures). In this work we do not include quantum tunneling; this issue is further discussed in Sect.\,\ref{ss:cmpros}. Also, we do not adopt the so called multilayer approach \citep[studied recently by e.g.][]{Garrod11, Taquet12} in which only the outermost surface layer is reactive. The exclusion of this process is also discussed in Sect.\,\ref{ss:cmpros}.

\subsection{Gas temperature calculation} \label{ss:gtcalc}

We have coupled radiative transfer calculations with a chemical model for a self--consistent determination of the gas temperature \citep[our procedure is similar to the calculations of][]{Bergin06}. In practice, when the MBES has been constructed \citep[using the method discussed in][]{Sipila11}, it is split into concentric spherical shells by linearly dividing the total radius of the core to pieces of equal length \citep[the same approach was followed also in][]{Sipila10}. Consequently, each shell has a unique density and a unique temperature. Chemical evolution is then calculated separately in each shell, and the abundances of all substances as a function of time are extracted (initially, we assume $T_{\rm gas} = T_{\rm dust}$).

\begin{figure}
\includegraphics[width=\columnwidth]{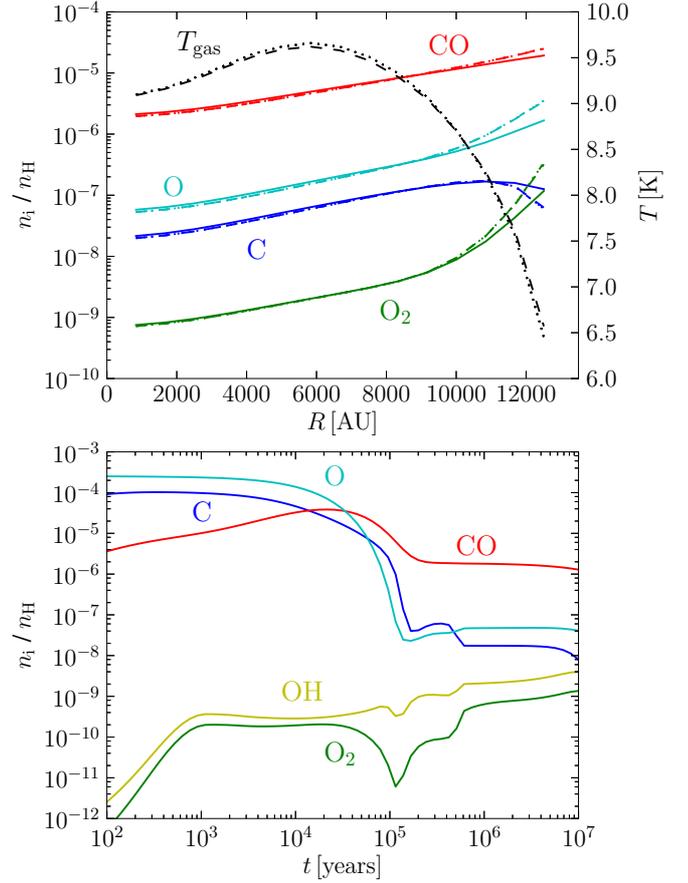}
\caption{{\sl Upper panel:} The abundances (with respect to $n_{\rm H}$, left--hand scale) of the cooling molecules (indicated in the figure) as a function of core radius in model B. Solid lines correspond to the initial chemistry calculation with $T_{\rm gas} = T_{\rm dust}$, other linestyles to subsequent iterations with $T_{\rm gas} \neq T_{\rm dust}$. The black curves represent $T_{\rm gas}$ at different iterations (right--hand scale). {\sl Lower panel:} The abundances of the cooling molecules and of OH as a function of time in a single-point chemical model with $n_{\rm H} = 2.0\times10^5$\,cm$^{-3}$ and $T_{\rm gas} = T_{\rm dust} = 9$\,K.}
\label{fig:titer}
\end{figure}

Using data from the chemistry model, molecular cooling calculations are performed with a Monte Carlo radiative transfer program \citep{Juvela97} -- the cooling calculations are similar to those presented in \citet{Juvela11}, i.e. we also employ the gas-grain coupling scheme of \citet{Goldsmith01}. In this study, we consider the cooling rates of six substances: O, O$_2$, C, $^{12}$CO, $^{13}$CO and C$^{18}$O. CO is expected to be the main coolant -- O and O$_2$ are included to study their possible contribution to the total cooling power (see Sect.\,\ref{ss:coolingrates}). Isotopes are not explicitly included in the chemical model -- when inputting CO radial abundance profiles into the radiative transfer program, we have assumed $^{12}$CO/$^{13}$CO and $^{12}$CO/C$^{18}$O abundance ratios of 60 and 500, respectively \citep[for typical values of the isotope ratios in the ISM, see e.g.][]{Wilson94}. We also assume that the dust grains are not significantly cooled or heated in interactions with the gas, so that $T_{\rm dust}$ stays constant throughout the calculation.

Molecular abundances change with time -- this means that the total cooling power may also change with time as molecules are depleted onto grain surfaces. To study how the gas temperature changes as a function of time owing to the changes in cooling power, we have run the cooling calculations using abundances corresponding to two different core ages ($t = 10^5$ and $t = 10^6$\,years). The radial abundances as a function of time are further discussed in Sect.\,\ref{s:results}.

After selecting the core age, the gas temperature as a function of core radius is calculated. To assure consistency, the (first approximation) $T_{\rm gas}$ profile is then fed back into the chemical model, which is now run assuming two separate temperature components. This yields new radial profiles for all substances, which are input to another round of cooling calculations. In practice only one or two iterations are needed -- the cooling molecules are insensitive to small ($\sim$1-2\,K) changes in temperature and hence the gas temperature converges very fast toward a single solution. To illustrate this, we plot in the upper panel Fig.\,\ref{fig:titer} the radial abundances of the cooling molecules at $t = 10^6$\,years in model B as a function of core radius. Evidently the change in radial abundances is very small -- after the initial chemistry calculation (solid lines) the molecular abundances change only at the edge of the core, and even there the change is negligible when further iterations are carried out (multiple curves with different linestyles practically superimposed). Also plotted in the figure are the gas temperatures at each iteration; the feedback from the minor changes in gas phase abundances to the gas temperature is negligible.

As an example of the time-evolution of the cooling molecules, we plot in the lower panel of Fig.\,\ref{fig:titer} the abundances of the cooling molecules in a single-point chemical model with total hydrogen density $n_{\rm H} = 2 \times 10^5$\,cm$^{-3}$ and temperature $T_{\rm dust} = T_{\rm gas} = 9\, \rm K$ -- these parameters represent roughly the center of model~B. At this density, the CO abundance is decreasing at $t = 10^5$ years due to depletion, but its abundance remains more or less constant from $t = 10^5$ to $t = 10^6$\,years. On the other hand, the abundances of C and O$_2$ change by many orders of magnitude due to depletion. The behavior of $\rm O_2$ around $t = 10^5$ years can be explained with the OH abundance. At earlier times, OH is primarily destroyed by N, C and O (in that order). After these atoms disappear from the gas phase, OH is destroyed mainly by $\rm H_3^+$ and $\rm N_2H^+$, but the destruction rate is an order of magnitude lower. Both before and after $t = 10^5$ years, OH is mainly produced in the dissociative electron recombination of $\rm H_3O^+$; the total production rate also decreases beyond $10^5$ years, but the production rate decreases less than the destruction rate, leading to an OH abundance increasing with time. $\rm O_2$ is mainly produced in the neutral exchange reaction
\begin{equation} \label{o2formation}
\rm O + OH \rightarrow O_2 + H \, 
\end{equation}
and mainly destroyed in
\begin{equation} \label{o2destruction}
\rm C + O_2 \rightarrow CO + O \, ;
\end{equation}
because O disappears from the gas slightly faster than C, reaction (\ref{o2destruction}) dominates over reaction (\ref{o2formation}) before $10^5$ years and the abundance of $\rm O_2$ decreases. When the O abundance settles to a (more or less) constant value, the increasing OH abundance helps to increase also the $\rm O_2$ abundance. 

Figure \ref{fig:titer} shows that the abundances of the cooling molecules can vary significantly with time. We shall see in Sect.\,\ref{s:results} that this time evolution is reflected in the total cooling power of the core and hence on $T_{\rm gas}$.

\subsection{Simulated spectra}

Owing to the time evolution of the molecular abundances, one can expect line emission radiation observed toward starless cores to also change with time. That is, the line profiles of main tracer molecules such as $\rm N_2H^+$ and $\rm ^{13}CO$ are likely to be different depending on the core age. To quantify the effect of temporal variation in molecular abundances on line emission radiation, we have calculated simulated line emission spectra using the radiative transfer program discussed above. We model the 1 -- 0 transitions of $\rm ^{13}CO$, $\rm N_2H^+$ and $\rm HCO^+$ as well as the (1,1) inversion transition of $\rm NH_3$. Molecular collision data is taken from the LAMDA database \citep{Schoier05}.

In these calculations we study the model B core, assuming a distance of 100\,pc. The beam FWHM is set equal to the core radius (126\arcsec \, at the assumed distance, cf. Table\,\ref{tab1}). We assume a turbulent linewidth of 0.22\,km\,s$^{-1}$. The results of these simulations are presented in Sect.\,\ref{ss:lineem}.

\section{Results} \label{s:results}

\subsection{Radial abundances of the cooling molecules} \label{ss:radcool}

\begin{figure*}
\includegraphics[width=17cm]{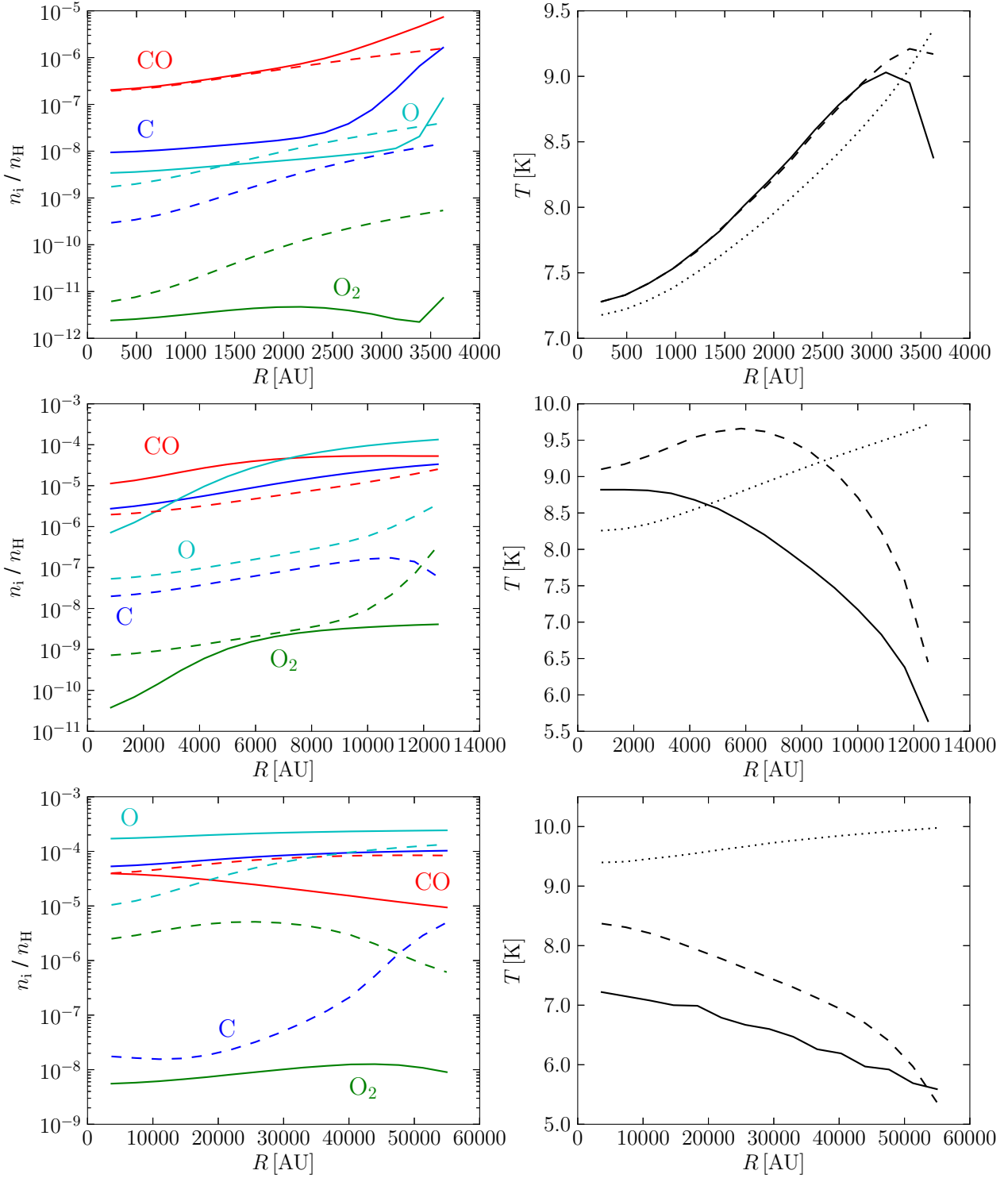}
\caption{{\sl Left-hand panels:} Radial abundances (with respect to $n_{\rm H}$) of the cooling molecules (indicated in the figure) in models A (top), B (middle) and C (bottom). Solid lines correspond to $t = 10^5$\,years, dashed lines to $t = 10^6$\,years. {\sl Right-hand panels:} The temperature profiles in models A to C. Solid lines correspond to $t = 10^5$\,years, dashed lines to $t = 10^6$\,years. Also plotted in each panel is the dust temperature (dotted lines).}
\label{fig:coolabu}
\end{figure*}

Molecular abundances change with time, and this affects in particular the total cooling power. We plot in the left--hand panels of Fig.\,\ref{fig:coolabu} the radial abundances of the cooling molecules at different times in models A, B and C. It is evident that the abundances are strongly dependent on both time and density. For example, atomic carbon is more abundant than atomic oxygen in model~A at $t = 10^5$\,years, but the situation is reversed as the core grows older. At lower densities (models~B and C), atomic carbon is generally less abundant than atomic oxygen at both displayed times. The difference between the two species is particularly large in model~C at $t = 10^6$\,years. As illustrated by the lower panel of Fig.\,\ref{fig:titer}, the abundances of C and O can change drastically (owing to depletion) in a relatively short time interval; this kind of behavior is accentuated in the low density of model~C where chemical evolution is slow in general. Given enough time, the abundances of C and O would settle to similar levels even in model~C.

\begin{figure*}
\centering
\includegraphics[width=17cm]{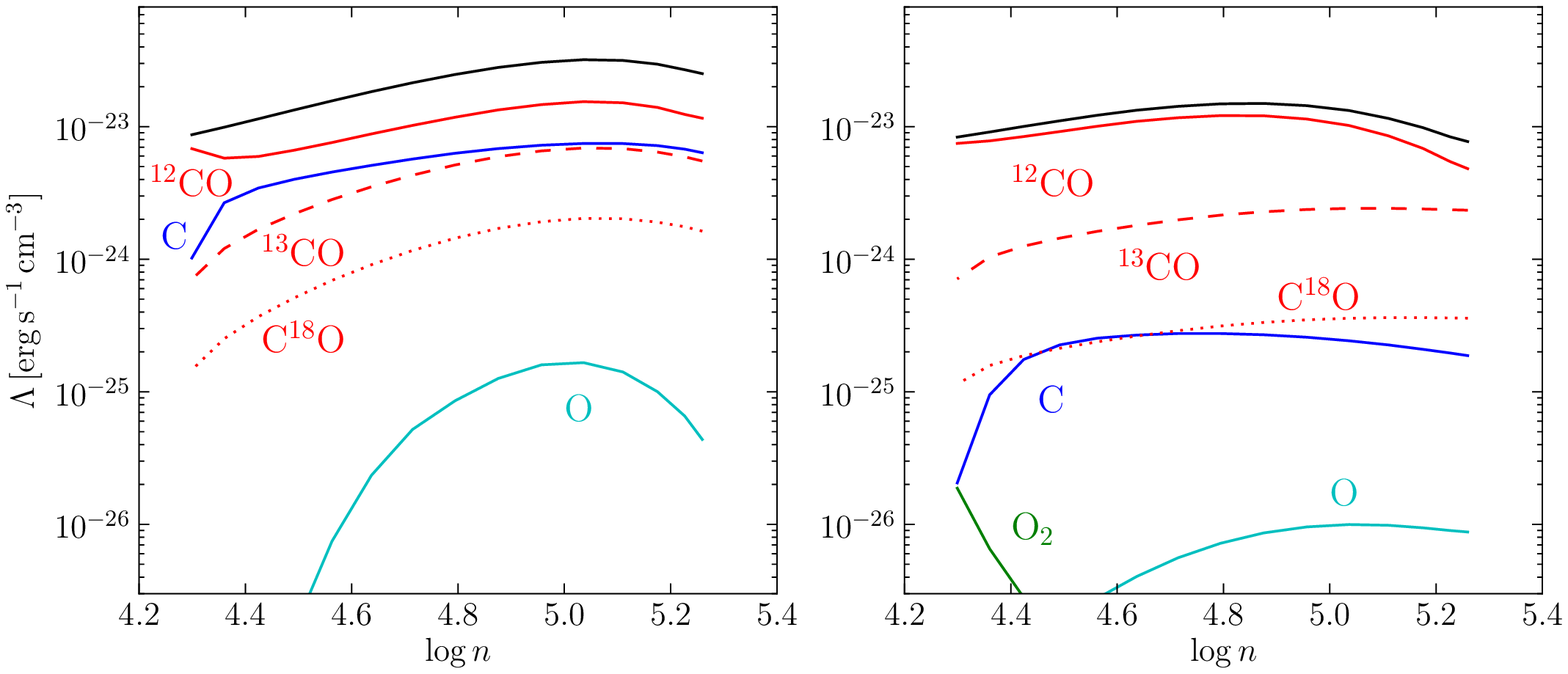}
\caption{The cooling rates of the cooling molecules (indicated in the figure) at $t = 10^5$\,years (left panel) and at $t = 10^6$\,years (right panel) in model~B as a function of gas density. Also plotted in both panels is the total cooling power (black lines).}
\label{fig:coolrates}
\end{figure*}

Molecular oxygen (green) evolves differently from atomic carbon and atomic oxygen; while its abundance decreases toward higher densities owing to depletion, the radial abundance of $\rm O_2$ is higher in an older core. At high densities, the growth of the radial abundance with time is about one order of magnitude, but up to two orders of magnitude at the low density of model~C. It should be noted that at low density, the $\rm O_2$ abundance evolves differently than in the lower panel of Fig.\,\ref{fig:titer}; the local minimum present in Fig.\,\ref{fig:titer} is, at low density, replaced by a local maximum. We note that our model~C predicts similar $\rm O_2$ abundances than the model of \citet{Hollenbach09}, even though they used the 800\,K value for the binding energy of atomic oxygen. Furthermore, the $\rm O_2$ abundances in model~C agree with the abundances observed by \citet{Goldsmith11}, although they observed regions with much higher temperatures ($\sim 100$\,K) than in our models.

CO (red) is fully depleted $t = 10^5$ years in model~A in the center of the core, but at the edge depletion is still underway. CO is fully depleted throughout the core by $t = 10^6$ years. In model~B at $t = 10^5$ years, CO depletion has just begun in the center, but at the edge CO has not yet had time to reach its maximum abundance (the evolution is similar to that shown in the lower panel of Fig.\,\ref{fig:titer}). Again at $t = 10^6$ years CO is fully depleted throughout the core. The inward-increasing CO gradient seen in model~C at  $t = 10^5$ years is due to slower chemical evolution at the edge. In this model there is very little CO depletion at $t = 10^6$ years even in the core center.

\subsection{Gas temperatures and cooling rates} \label{ss:coolingrates}

The abundances of the cooling molecules vary considerably with time; the contributions of different molecules to the core cooling can change accordingly. Also, the total cooling power may decrease as molecules are depleted onto grain surfaces. The right--hand panels of Fig.\,\ref{fig:coolabu} plot the gas temperature in models A to C at two different times; evidently, the gas temperature changes with time. The changes in $T_{\rm gas}$ as a function of time are the smallest in model~A. The core is warmer at the edge at $t = 10^6$ years owing to depletion; the main coolant CO is not yet fully depleted at the edge at $t = 10^5$ years. In models B and C the time variation of $T_{\rm gas}$ is more evident; however, in both cases the $T_{\rm gas}$ time variation in a given point of the core is of the order of 1\,K; these results agree with the calculations of \citet{Bergin06} who also reported temporal $T_{\rm gas}$ variations of $\sim$1\,K (albeit in a different scenario).

To quantify the total cooling power and the contributions of the various molecules to it, we plot in Fig.\,\ref{fig:coolrates} the cooling rates of the individual cooling molecules (indicated in the figure) and the total cooling rate as a function of gas density in model~B at $t = 10^5$ and $t = 10^6$ years. There are clear differences between the two time steps. Atomic carbon is the second most important coolant at $t = 10^5$ years, but rather insignificant at $t = 10^6$ years. The cooling rate of $\rm O_2$ is negligible and shows up only at the core edge at $t = 10^6$ years due to an increased abundance (cf. Fig.\,\ref{fig:coolabu}). At this time, atomic carbon is significantly depleted which translates to a low cooling rate. Atomic oxygen is of little significance in terms of cooling power. $^{12}\rm CO$ is the main coolant in both cases; its contribution totally dominates the total cooling power at later times. However, because of depletion, the abundance of CO drops by about an order of magnitude from $t = 10^5$ to $t = 10^6$ years (cf. Fig.\,\ref{fig:coolabu}). Accordingly, the total cooling rate is lower at the later time step and the core warms up as a result; a similar effect is seen in models A and C (Fig.\,\ref{fig:coolabu}). 

\citet{Goldsmith01} argued that O and $\rm O_2$ would be insignificant coolants in dark clouds due to their low abundance. Our models confirm this statement. From our results it is also evident that the individual contributions of the various molecules to the core cooling evolve with time.

\subsection{Tracer molecules} \label{ss:tracers}

\begin{figure*}
\centering
\includegraphics[width=17cm]{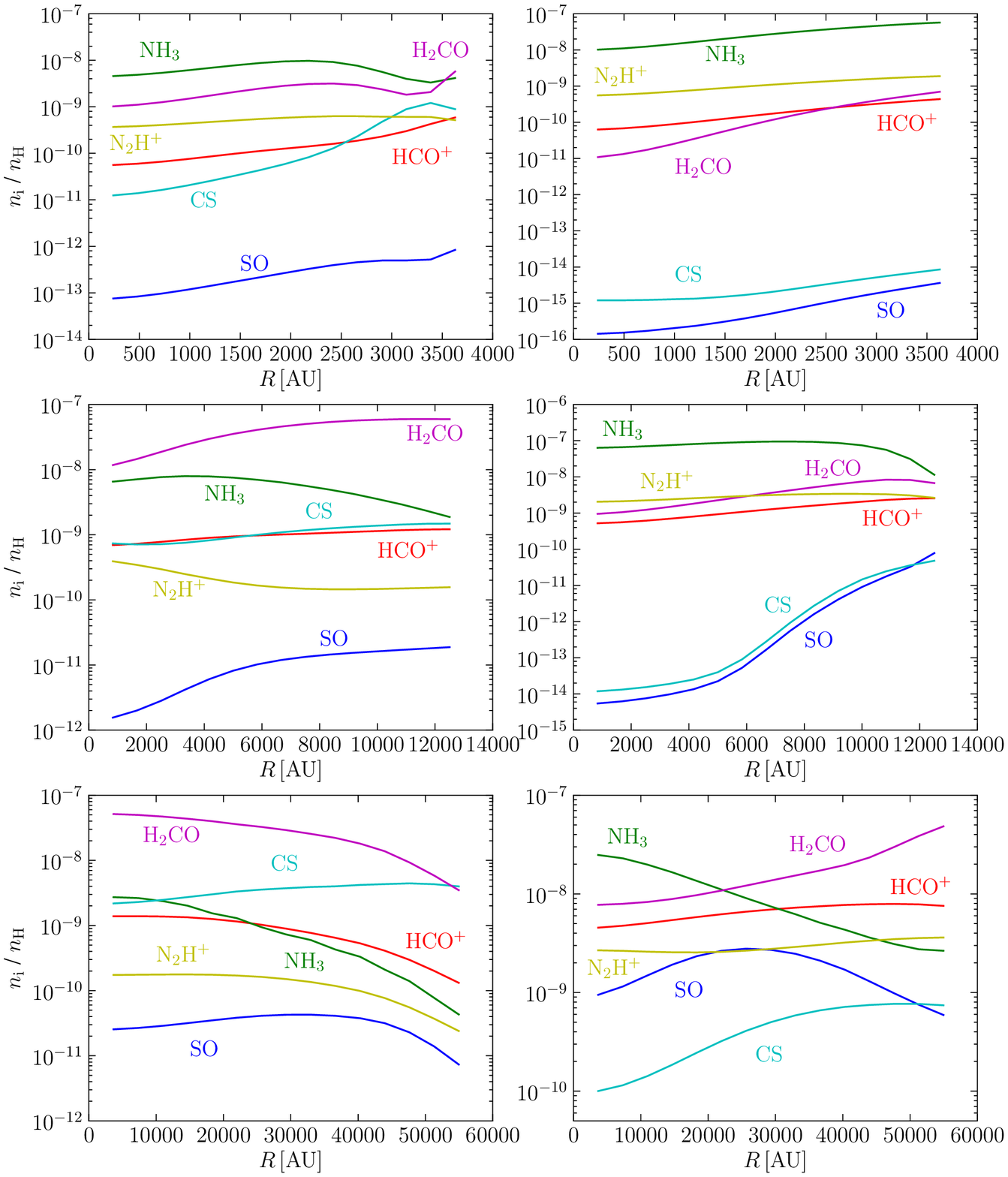}
\caption{Radial abundances (with respect to $n_{\rm H}$) of common tracer molecules (indicated in the figure) in models A (top), B (middle) and C (bottom) at $t = 10^5$\,years (left--hand panels) and $t = 10^6$\,years (right--hand panels).}
\label{fig:tracerabu}
\end{figure*}

Starless cores are often observed in various transitions of a multitude of molecules. This approach makes it possible to probe areas of different densities. However, as demonstrated by the cooling molecules above, the abundance of a given molecule can vary strongly with core radius, and most molecules cannot be used to probe the central parts of the cores because of depletion onto grain surfaces. Since the chemistry evolves relatively slowly in starless cores, we expect to see different abundance gradients, and particularly different degrees of depletion, in cores of different ages.

\begin{figure*}
\includegraphics[width=17cm]{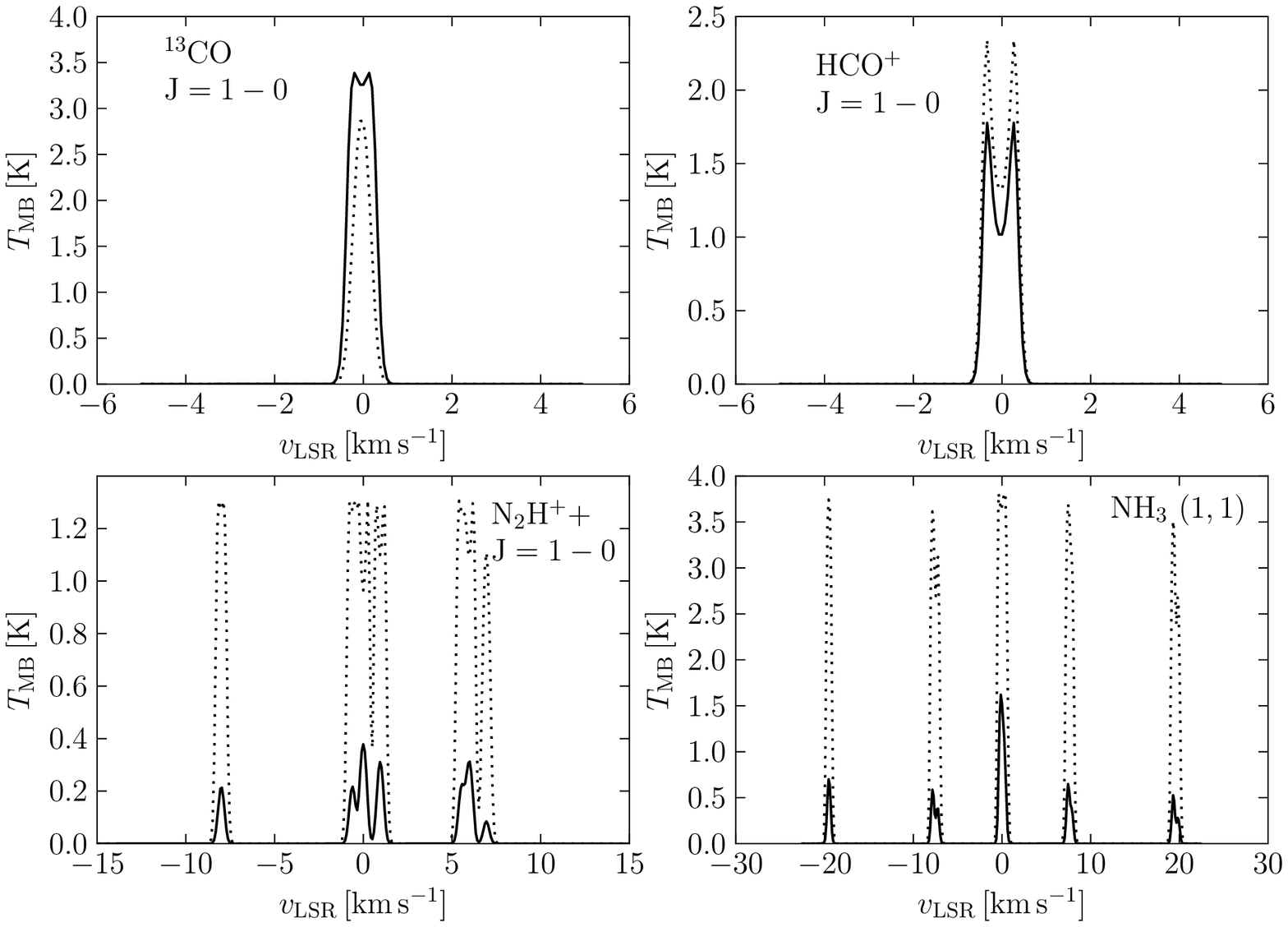}
\caption{Simulated line emission profiles (molecules and transitions indicated in figure) in model~B assuming a distance of 100\,pc. Solid and dotted lines correspond to core ages of $t = 10^5$ years and $t = 10^6$ years, respectively. In the simulations, the beam width is set equal to the core radius (126$\arcsec$ at the assumed distance.)}
\label{fig:emission}
\end{figure*}

\begin{figure*}
\centering
\includegraphics[width=17cm]{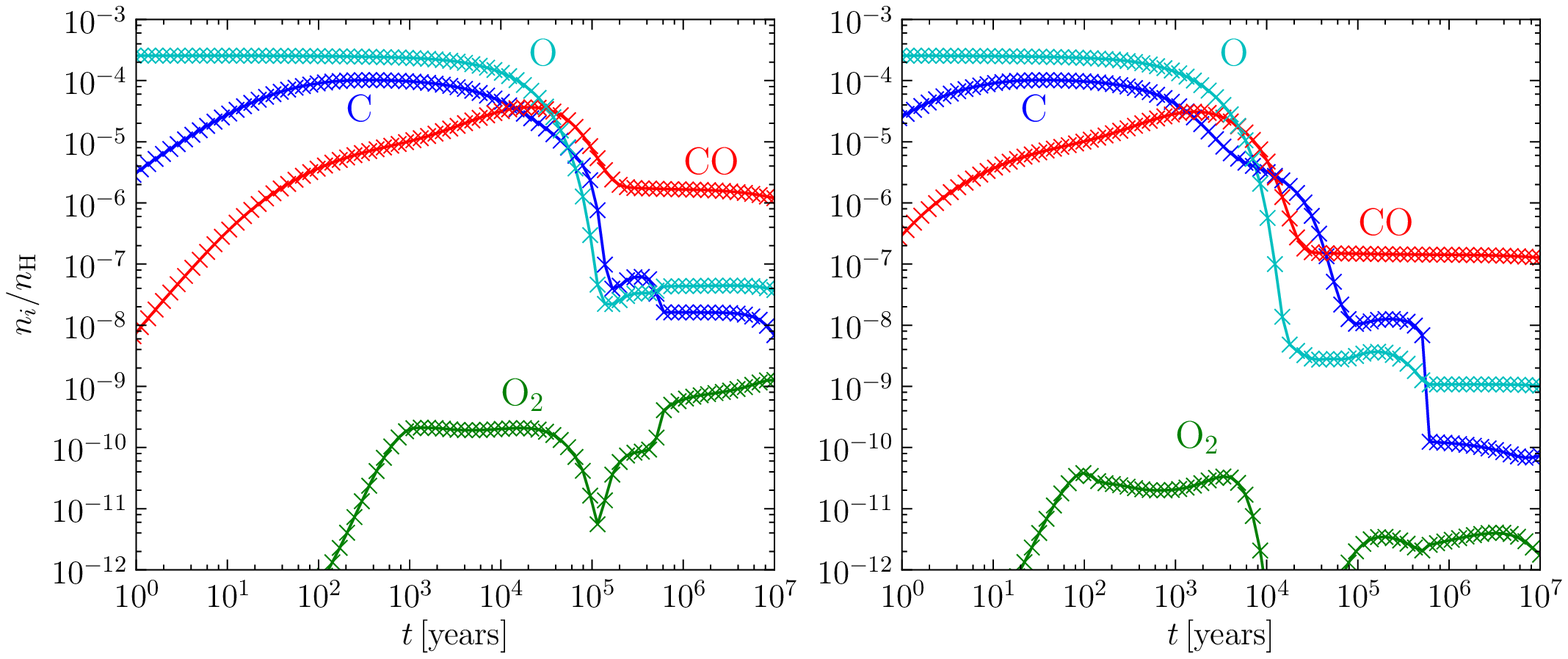}
\caption{The abundances (with respect to $n_{\rm H}$) of the cooling molecules (indicated in the figure) as a function of time in single-point chemical model with $n_{\rm H} = 2 \times 10^5$\,cm$^{-3}$ (left-hand panel) and $n_{\rm H} = 2\times10^6$\,cm$^{-3}$ (right-hand panel). Solid lines: H tunneling included; crosses: no tunneling.}
\label{fig:quantumtest}
\end{figure*}

To demonstrate this, we plot in Fig.\,\ref{fig:tracerabu} the radial abundance profiles of $\rm NH_3$, $\rm N_2H^+$, CS, SO, $\rm HCO^+$ and $\rm H_2CO$ for two different core ages in models A, B and C. It is evident that, in most cases, abundances change by about one order of magnitude when comparing the core center and edge, although the abundance gradients get somewhat less steep as the density increases. Notable exceptions are the sulfur-bearing molecules, the chemistry of which is however not very well understood due to a lack of experimental data on rate coefficients. Core age is a very important factor. For example, in all models the abundances of $\rm NH_3$ and $\rm N_2H^+$ are about an order of magnitude {\sl higher} throughout the core at $t = 10^6$ years than at $t = 10^5$ years. $\rm HCO^+$ displays similar behavior at low density, but at higher densities its radial abundance hardly changes with time. In contrast, the abundances of formaldehyde, SO and CS tend to drop as the cores get older. In models A and B, the sulfur-bearing molecules CS and SO are highly depleted at $t = 10^6$ years.

Similarly to the cooling molecules, in model~C the abundances are rather undeveloped even at $t = 10^6$ years due to slow chemical evolution. In this model the density is so low (cf. Table\,\ref{tab1}) that none of the species is significantly depleted even at $t = 10^6$ years. Indeed, model~C is characterized by generally inward--increasing abundance gradients at $t = 10^5$ years.

\subsection{Line emission profiles} \label{ss:lineem}

The temporal variation of radial molecular abundances means from an observational point of view that observed line emission profiles depend on the core age, and hence line emission observations can, at least in principle, be used in conjunction with theoretical models to (roughly) estimate the age of a starless core.

To demonstrate how the changes in radial abundances as a function of time translate to line emission radiation, we plot in Fig.\,\ref{fig:emission} simulated line profiles of the 1 -- 0 transitions of $\rm ^{13}CO$, $\rm N_2H^+$ and $\rm HCO^+$ and the (1,1) inversion transition of $\rm NH_3$ in model B for two different core ages. The $\rm ^{13}CO$ profile is not significantly altered as the core grows older, although the peak intensity and optical depth decrease somewhat due to molecular depletion. The $\rm HCO^+$ profiles present significant self-absorption; the optical depths at $t = 10^5$ and $t = 10^6$ years are 12.5 and 15.1, respectively. The increased optical depth at later times is due to a slightly larger abundance at the core edge (Fig.\,\ref{fig:tracerabu}). The $\rm N_2H^+$ and $\rm NH_3$ line profiles are significantly different depending on assumed core age; this is again due to increased radial abundances (Fig.\,\ref{fig:tracerabu}).

Figure \ref{fig:emission} shows that there can be significant variation in line emission depending on the age of the studied core. For typical starless core densities ($10^4 - 10^5$\,cm$^{-3}$), $\rm NH_3$ and $\rm N_2H^+$ can serve as useful tracers for cores as old as $t = 10^6$ years. At even higher densities, represented here by model A, $\rm NH_3$ is depleted toward the core center but $\rm N_2H^+$ still retains a more or less constant radial abundance (Fig.\,\ref{fig:tracerabu}), which supports the notion that $\rm N_2H^+$ is a viable tracer of high--density regions \citep[e.g.][]{Tafalla02, Pagani07}.

Finally, we note that in the radiative transfer calculations, we have assumed zero infall velocity and that the core does not rotate. This means that the line profiles presented in Fig.\,\ref{fig:emission} represent an idealized case of a perfectly static core. As real cores are often either oscillating or contracting \citep[e.g.][]{Broderick08, Lee11}, care has to be taken when comparing simulated line profiles with observations. For example, the $\rm N_2H^+$ and $\rm NH_3$ profiles at $t = 10^6$ years shown in Fig.\,\ref{fig:emission} are optically thick; the high optical thicknesses (8.12 and 7.63 for $\rm N_2H^+$ and $\rm NH_3$, respectively) are reduced to 6.06 and 6.61, respectively, if one assumes an infall velocity equal to the turbulent velocity used here (0.22\,km\,s$^{-1}$), producing sharper and slightly more intense line profiles.

\section{Discussion} \label{s:discussion}

In this section, we discuss some aspects of chemical modeling that are not included in our calculations. We also discuss molecular abundances and the associated depletion predicted by our models. We end the section with a discussion on the issues involved in determining gas temperatures and its possible impact on e.g. core stability.

\subsection{Quantum tunneling and multilayer grain chemistry} \label{ss:cmpros}

We investigated the effect of quantum tunneling of atomic hydrogen on gas phase abundances by running multiple single-point chemical models representing a range of densities, with H tunneling either included or excluded. In these tests, we set $T_{\rm dust} = T_{\rm gas} = 10$\,K. Following \citet{HHL92}, whenever H is one of the reactants of a given surface reaction the $\kappa$ factor is calculated as
\begin{equation}
\kappa = \exp \left[ -2 \, ( a / \hbar) \, (2 \, \mu \, E_a)^{1/2} \right] \, ,
\end{equation}
where $a$ is an arbitrary barrier width and $\mu$ the reduced mass. \citet{HHL92} assumed $a = 1\,\AA$; we use this value as well. The value of $a$ is a source of uncertainty in chemical models -- recently, for example, \citet{Garrod11} used a value $a = 2\,\AA$ based on results by \citet{Andersson09} and \citet{Goumans10}. The diffusion rate of a tunneling reactant  (in this case H) is calculated as
\begin{equation}
R_{\rm diff}^q = {\nu_0 \over N_s} \exp \left[-2 \, ( a / \hbar) \, (2 \, m \, E_{\rm diff})^{1/2} \right]\, .
\end{equation}
Figure \ref{fig:quantumtest} shows comparisons of the abundances of the cooling molecules as a function of time in a single-point chemical model with densities $n_{\rm H} = 2\times10^5$\,cm$^{-3}$ (left-hand panel) and $n_{\rm H} = 2\times10^6$\,cm$^{-3}$ (right-hand panel). Evidently, the abundances of the cooling molecules are virtually unaffected by the inclusion of H tunneling. However, the inclusion of tunneling does affect, e.g., the abundances of hydrocarbons on grain surfaces -- this is to be expected, since tunneling decreases the hydrogenation timescale on grain surfaces as hydrogen atoms scan the grain surfaces faster. A more extensive study of the effect of quantum tunneling is beyond the scope of this paper and it is unclear if including tunneling of, e.g., C and O would affect the abundances of the cooling molecules. We have performed the calculations presented in this paper without tunneling in order not to introduce many new unknown factors into the modeling. However, our tests indicate that at least in the case of H tunneling the results of this paper should remain unaffected. Also, we do not expect including the so-called ``modified rate equation method'' \citep[e.g.][]{Garrod08, Du11} in the modeling of grain surface reactions to influence our results, as we consider rather large (0.1\,$\mu$m) grains in a low-temperature environment.

As mentioned in Sect.\,\ref{ss:chemmodel}, we do not consider in this paper the so-called multilayer approach, where only the outermost grain surface layer is reactive and the mantle under the surface is chemically inert. As pointed out by \citet{Taquet12}, this means in particular that models such as the one considered here may underestimate the abundances of radicals on grain surfaces. It is unclear to what extent this affects gas phase abundances. There may be large differences between the multilayer and one-layer models at higher densities, where the depletion timescale is shorter. A quantitative study of this issue is certainly warranted, but beyond the scope of this paper.

\subsection{Molecular abundances and depletion} \label{ss:molabus}

We have constructed chemical models of starless cores assuming a modified Bonnor-Ebert sphere for the physical structure of the cores. All of the core models considered here present a density gradient of about one order of magnitude; this property is imposed on the model cores due to the requirement of stability. To study larger density gradients, one could in principle consider supercritical core models with the non--dimensional critical radius above $\xi \sim 6.5$. However, in the context of supercritical and hence unstable cores it is probably not feasible to study model cores with ages much above $\sim 10^5$ years, since the core lifetime is in this case expected to be very short compared to typical chemical timescales.

In the core models considered here, we typically find molecular depletion of about one order of magnitude across the cores\footnote{This statement means that a given species is more depleted at the core center than at the edge by an order of magnitude. For example in model~A at $t = 10^6$ years, CO is depleted by $\sim$ 3 orders of magnitude at the core center and by $\sim$ 2 orders of magnitude at the edge. Radial abundance profiles such as those presented here do not trace the ``absolute'' amount of depletion, but reflect the differential depletion arising from differences in density between different parts of the core.}. The depletion behavior of different species is however non-trivial: some species present flat radial abundance profiles while others are heavily depleted toward the core center. As demonstrated by Fig.\,\ref{fig:tracerabu}, inward-increasing abundance gradients are possible even in old ($t \sim 10^6$~years) cores, as long as the density is low. The density of the core plays an important role: abundance gradients are generally rather monotonous in high-density cores, while low-density cores can present more exotic abundance gradients because of the long timescales.

In this study we have not attempted to identify viable tracer species; the abundance profiles presented in Sect.\,\ref{ss:tracers} serve to demonstrate possible changes as a function of time that one could expect to observe in starless cores. In principle, models such as those presented here could be used for identifying tracer candidates, but this kind of analysis should be carried out using a wide parameter space, instead of fixing most physical parameters such as grain size and cosmic ray ionization rate, as was done here.

As noted in Sect.\,\ref{ss:chemmodel}, in all models the gas phase abundances are initially atomic, with the exception of hydrogen which is almost totally in $\rm H_2$. In this paper we have studied dense, deeply embedded cores -- it is not clear to what extent the assumption of atomic initial abundances is applicable in this case. Starting from molecular instead of atomic abundances, i.e. assuming chemical evolution previous to the starless core phase, will change the chemical evolution timescales. To study the possible effect of the initial abundances on the results of this paper, we have run multiple single-point chemical models assuming either atomic or molecular initial abundances. The molecular initial abundances were obtained by computing chemical evolution in a low density ($\sim$10$^3$\,cm$^{-3}$) model; the so-obtained abundances were then used as input for higher density models. The atomic and molecular cases predict identical abundances after a time comparable to the age of the progenitor diffuse cloud -- that is, if we let the diffuse model evolve for e.g. $10^6$ years before extracting the abundances (i.e. initial abundances for the high density model), then the high density model predicts identical abundances at $t \gtrsim 10^6$ years regardless of the assumed initial abundances. Thus, the results of this paper should hold to good accuracy whether the initial abundances are atomic or molecular, provided that the progenitor diffuse core is not very old. We have refrained from more extensive testing of this issue because the present model does not allow for self-consistent merging of the diffuse and higher density models (by, e.g., allowing the diffuse cloud to gravitationally collapse into a denser configuration); certainly the statements made above should not be regarded as a general conclusion. For example in the context of collapse models, it has been shown that nitrogen chemistry is sensitive to the initial conditions \citep[][]{Flower06}.

Recently, \citet{Hily-Blant10} and \citet{Padovani11} have studied observationally \citep[and by using chemical models;][]{Hily-Blant10} the abundances of CN, HCN and HNC in starless cores. They find that both HNC and HCN are present in appreciable abundances in the gas phase at densities where CO is depleted, and that the HNC/HCN ratio is close to unity. Our models reproduce both of these results. The HNC/HCN ratio depends on time; in young cores HNC is more abundant, but the situation is reversed as the core grows older. However, the ratio stays very close to unity at all times. For the abundances of CN, HCN and HNC we obtain values similar (within an order of magnitude) to the theoretical results of \citet{Hily-Blant10}, although they consider a collapse model whereas our model is static. In conclusion, our models also fail to explain the observations of \citet{Hily-Blant10} and \citet{Padovani11} -- the possible reasons for this failure are discussed in \citet{Hily-Blant10}.

\subsection{The $\rm NH_3 \, /  \, \rm N_2H+$ abundance ratio}

The $\rm NH_3$/$\rm N_2H^+$ abundance ratio has been studied in several cores and observed to increase toward the core centers \citep[e.g.][]{Tafalla02, Hotzel04}. In these studies, the derived core densities are in the range $10^4 - 10^5$\,cm$^{-3}$, corresponding roughly to our model~B. Comparison between these observations and model~B yields a rather good agreement at $t = 10^5$\,years, our models reproducing rather well the radial abundances and consequently the $\rm NH_3$/$\rm N_2H^+$ abundance ratio, even though we do not attempt detailed modeling of any of the observed cores. Our models indicate that at high densities ($> 10^5$\,cm$^{-3}$) the $\rm NH_3$/$\rm N_2H^+$ abundance ratio starts to fall as $\rm NH_3$ begins to deplete (while $\rm N_2H^+$ is still relatively unaffected). This is an interesting result and warrants observational verification.

Similar modeling results of the $\rm NH_3$/$\rm N_2H^+$ abundance ratio at intermediate densities have been published by \citet{Aikawa05}. In their models, however, $\rm N_2H^+$ depletes efficiently at higher densities, leading to a larger $\rm NH_3$/$\rm N_2H^+$ abundance ratio than in our models. The reason for this discrepancy is probably that \citet{Aikawa05} included in their reaction set the branching ratios of \citet{Geppert04} for the $\rm N_2H^+ + e^-$ recombination reaction -- these branching ratios have since been refuted and are not included in the {\tt osu\_03\_2008} reaction file. In the present model, the recombination yields $\rm N_2 + \rm H$ (90\,\%) and $\rm NH + \rm N$ (10\,\%).

\citet{Fontani12} have recently studied observationally the $\rm NH_3$/$\rm N_2H^+$ ratio in prestellar core candidates and found that the ratio increases strongly toward higher densities; in their Appendix~A, they also show using a gas phase chemical model that the $\rm NH_3$/$\rm N_2H^+$ ratio should increase with increasing depletion. Our model seemingly predicts the opposite: the ratio decreases at high density (corresponding to a higher degree of depletion). However, comparison of the results predicted by our model and that of \citet{Fontani12} is difficult, because their model does not include a time-dependent treatment of freeze-out; instead, they set a value for the depletion factor $f_{\rm D}$ describing a uniform depletion of species heavier than helium. In our model, different species present highly variable degrees of depletion. This is demonstrated by e.g. the lower panel of Fig.\,\ref{fig:titer}, where CO is depleted by less than 2 orders of magnitude at $t = 10^6$\,years, while e.g. C, O and N (not shown in the figure) are depleted by $\sim$ 3 orders of magnitude. This differential depletion depends both on density and on the assumed core age -- one cannot in our model assign a constant describing a general degree of depletion, and consequently our results cannot be directly compared with those of \citet{Fontani12}. In our model, the adsorption of $\rm NH_3$ increases in importance as the density increases; this effect is primarily responsible for the decreasing $\rm NH_3$/$\rm N_2H^+$ ratio at high density.

Finally, we note that our chemical model reproduces the modeling results of \citet[][lower panel in their Fig. A-1]{Fontani12} if we ``turn off'' the gas-grain interaction, i.e. if we consider gas phase chemistry and vary the metal abundances, although the dependence of the $\rm NH_3$/$\rm N_2H^+$ ratio on the depletion factor is in our model somewhat less steep than in \citet{Fontani12}.

\subsection{Gas temperature}

The gas temperature is a function of time. As a core grows older, the total core cooling power decreases as cooling species (most importantly CO) are depleted onto grain surfaces and as a result, the gas temperature rises. However, in a typical core lifetime, this change may be only $\sim 1-2$\,K depending on the gas density \citep[see also][]{Bergin06}. Test calculations indicate that the feedback of small changes in gas temperature to the chemistry is small. It should be noted that this effect is not self-consistently modeled here as the gas temperature is not allowed to evolve in time; we have determined the gas temperature at different times using the gas phase abundances appropriate to those times. A more complete model taking into account dynamical changes in both the molecular abundances and the gas temperature could of course be built, but this approach is computationally much more demanding than the approach considered here and would probably not change our results significantly (as suggested by the rather small feedback of $T_{\rm gas}$ on the chemistry).

Gas temperature calculations are often carried out by either assuming constant radial abundances or assuming an arbitrary degree of depletion (using e.g. a step function) toward the core center that is the same for all cooling species \citep[e.g.][]{Keto05, Juvela11}, although calculations taking into account the chemistry have also been performed \citep[e.g.][]{Keto08, Keto10}. Our results indicate that in reality the abundances are probably non-trivial; species deplete at different rates and the degree of depletion varies from species to species. Furthermore, depletion tends to increase radially toward the core center. Overestimating the degree of depletion leads to a warmer core, while underestimating it leads to a colder core (see Sect.\,\ref{ss:coolingrates}). Figures \ref{fig:coolabu} and \ref{fig:tracerabu} indicate that abundances in dense cores with advanced ages can be described by a (nearly) uniform degree of depletion, but this approximation breaks down for either young dense cores or cores with low gas densities. We have not made a direct comparison between our models and one assuming e.g. constant radial profiles with arbitrary depletion near the center, but the differences in gas temperature predicted by the two approaches may not be larger than a few K -- this of course also depends on the assumed cooling molecules.

As a rule, the gas temperature in our models drops toward the core edge. This is because at the core edge the gas-dust coupling is weaker and the photon escape probability higher than at the center \citep[see also][]{Juvela11}. We assume $\rm A_V = 10$ at the edge, which means that heating of the cores by both the photoelectric effect and external UV radiation is negligible. In our models, there is no low-density gas outside the core that could provide heating by, e.g., diffusion. The presence of an external gas component would probably raise the outer core temperature somewhat; the chemistry would probably not be greatly affected (as discussed above), but a higher temperature could produce stronger line emission for the substances that are abundant in the outer, lower density regions (such as CO).

In the calculations presented here, the physical core model is determined by the dust temperature and remains unchanged through subsequent chemistry and gas temperature calculations. This should be a good approximation in low mass ($ < 1$\,$M_\odot$) cores, but at higher core masses the lower temperature of the gas in areas of moderate $\rm A_V$ (as opposed to that of the dust; see Fig.\,\ref{fig:coolabu}) may modify the density profile and affect the stability of the core \citep{Bergin06, Sipila11}, although the presence of turbulent pressure (neglected here) could counter this. Furthermore, at low values of $\rm A_V$, the gas heats up again due to photoelectric heating \citep[e.g.][]{Bergin06, Keto10}, providing additional support. Of course, changes in the density profile may also alter the chemistry, affecting the core cooling. Modeling these effects is beyond the scope of this paper, but a study of core stability including chemistry \citep[expanding the analysis of][]{Sipila11} is planned.

\section{Conclusions} \label{s:conclusions}

We have calculated radial molecular abundances and gas temperatures in core models representing starless cores in a semi-self-consistent way by inputting abundance profiles from a chemical model into a radiative transfer program to determine the gas temperature. The gas temperatures were calculated at different timesteps corresponding to different stages of chemical evolution. The gas temperature changes with time; the cores warm up slightly as cooling molecules are depleted onto grain surfaces. The interplay of evolving chemical abundances and the induced changes in gas temperature can translate into significant changes in line emission radiation as the core grows older.

We examined the simulated radial abundances of some common tracer molecules as a function of time in multiple core models covering a range of densities. The models support earlier results in the literature that $\rm NH_3$ and $\rm N_2H^+$ can resist depletion even at densities of $10^5 - 10^6$\,cm$^{-3}$. On the other hand, many species present inward-increasing abundance gradients at low densities where the depletion timescale is long. In low density regions, chemical evolution is slow and significant depletion may not occur in a typical core lifetime. Molecular abundance gradients are in many cases non-trivial, especially for young cores with low densities.

One interesting aspect of starless core chemistry, deuteration, is not covered in this paper. Deuteration is expected to be significant in the dense centers of the cores where heavy molecules are depleted onto grain surfaces and thus the deuteration process can proceed relatively unhindered. Because strong deuteration is effectively confined into a relatively small area in the core center, this process is not likely to have a notable influence on the gas temperature and hence on the results of this paper. However, light deuterated molecules such as $\rm H_2D^+$ and $\rm D_2H^+$ are important tracers of high-density gas. Because of its importance in high-density regions, inclusion of deuteration to the chemistry model is underway.

\acknowledgements

I thank the referee Dr. Paola Caselli for a thorough report which improved the paper. I also thank J. Harju for useful comments and suggestions on a draft of this paper and M. Juvela for technical assistance with the radiative transfer code. This work is supported by the Academy of Finland through grant 132291.

\bibliographystyle{aa}
\bibliography{19083.bib}

\end{document}